\documentclass{article}

\usepackage{amsmath}
\usepackage{PRIMEarxiv}

\usepackage[utf8]{inputenc} % allow utf-8 input
\usepackage[T1]{fontenc}    % use 8-bit T1 fonts
\usepackage{hyperref}       % hyperlinks
\usepackage{url}            % simple URL typesetting
\usepackage{booktabs}       % professional-quality tables
\usepackage{amsfonts}       % blackboard math symbols
\usepackage{nicefrac}       % compact symbols for 1/2, etc.
\usepackage{microtype}      % microtypography
\usepackage{lipsum}
\usepackage{fancyhdr}       % header
\usepackage{graphicx}       % graphics
\graphicspath{{media/}}     % organize your images and other figures under media/ folder
\usepackage{svg}
\usepackage[font=small,labelfont=bf]{caption}
\usepackage{float}
\usepackage{cite}
\usepackage[misc]{ifsym}
\usepackage{multirow}

\usepackage{xcolor}

\title{Network higher-order structure dismantling}

\author{
  Peng Peng\footnote{Institute of Fundamental and Frontier Studies, University of Electronic Science and Technology of China, 611731 Chengdu, P.R. China} 
  \and
  Tianlong Fan\footnote{School of Cyber Science and Technology, University of Science and Technology of China, 230026 Hefei, P.R. China\\ \hbox to 1.8em{\hss \textsuperscript{\Letter}}Corresponding authors: tianlong.fan@ustc.edu.cn, linyuan.lv@ustc.edu.cn}{\hspace{1mm}}
  \textsuperscript{\Letter}
  \and
  Linyuan L\"{u}
  \footnotemark[1]\textsuperscript{\hspace{0.5mm}}
  \footnotemark[2]\textsuperscript{\hspace{1mm}}
  \textsuperscript{\Letter}
}
\date{\today}

\begin{document}
\renewcommand{\thefootnote}{\arabic{footnote}}
\maketitle

\begin{abstract}
Diverse higher-order structures, foundational for supporting a network's ``meta-functions’’, play a vital role in structure, functionality, and the emergence of complex dynamics. Nevertheless, the problem of dismantling them has been consistently overlooked. In this paper, we introduce the concept of dismantling higher-order structures, with the objective of disrupting not only network connectivity but also eradicating all higher-order structures in each branch, thereby ensuring thorough functional paralysis. Given the diversity and unknown specifics of higher-order structures, identifying and targeting them individually is not practical or even feasible. Fortunately, their close association with $k$-cores arises from their internal high connectivity. Thus, we transform higher-order structure measurement into measurements on $k$-cores with corresponding orders. Furthermore, we propose the Belief Propagation-guided High-order Dismantling (BPDH) algorithm, minimizing dismantling costs while achieving maximal disruption to connectivity and higher-order structures, ultimately converting the network into a forest. BPDH exhibits the explosive vulnerability of network higher-order structures, counterintuitively showcasing decreasing dismantling costs with increasing structural complexity. Our findings offer a novel approach for dismantling malignant networks, emphasizing the substantial challenges inherent in safeguarding against such malicious attacks.
%179 words

\end{abstract}

\section{Introduction}

Given its profound implications in diverse dynamics and optimization problems \cite{albert2000error,moroneInfluenceMaximizationComplex2015,lu2013link,clusella2016immunization,lu2016vital}, the network dismantling \cite{cohen2001breakdown,braunsteinNetworkDismantling2016} persists as a focal point in the realm of network science, commanding substantial scholarly attention. For networks consisting of nodes connected through edges, a prevailing assumption suggests that the structural connectivity of the network stands as a prerequisite for its normal dynamics and functioning \cite{braunsteinNetworkDismantling2016}. As a result, disrupting this connectivity becomes a pivotal pursuit, with the aim of impairing network functionality or destabilizing the intricate dynamics stemming from interconnectivity. In particular, evaluating how network connectivity and functional states respond to network dismantling or attack behaviors can be accomplished by monitoring the condition of the giant connected component (GCC) \cite{schneiderMitigationMaliciousAttacks2011}. Additionally, this monitoring facilitates the assessment of the effectiveness and efficiency of dismantling strategies \cite{renGeneralizedNetworkDismantling2019}. Let us refer to this methodology as conventional network dismantling.

However, rapidly accumulating research indicates that the paradigm of conventional network dismantling is overly simplified, rendering it inadequate to explain certain common scenarios. For instance, complex functional behaviors can still emerge in many networks with relatively small scales \cite{krebs2002mapping,yan2017network}. Conversely, even when a large network is dismantled into smaller components, its functionality can still be maintained \cite{lehar2008high,moroneKcorePredictorStructural2019}. This is primarily because conventional dismantling often focuses on disrupting the network's overall connectivity. However, local clusters with high internal connectivity are frequently only mildly affected, especially with certain dismantling methods based on bridge edges \cite{wu2018bridges}, weak nodes \cite{moroneInfluenceMaximizationComplex2015}, and communities \cite{requiao2015fast, musciotto2023exploring}, as well as reinsertion-based operations \cite{pengUnveilingExplosiveVulnerability2023}. These methods aim to remove these local clusters as a whole to rapidly reduce the size of the GCC, driven by cost considerations. These diverse local clusters often serve as the structural foundation for high-order interactions, playing a crucial role in supporting local dynamics and the ``meta-functions’’ \cite{lambiotteNetworksOptimalHigherorder2019,battistonNetworksPairwiseInteractions2020}. Even when detached from the GCC, their functionality may still be maintained.

Across the fields of biology \cite{sanchez2019high}, neuroscience \cite{yuHigherOrderInteractionsCharacterized2011}, social systems \cite{alvarez2021evolutionary}, ecology \cite{grilliHigherorderInteractionsStabilize2017}, and engineering \cite{myersSoftwareSystemsComplex2003}, high-order interactions are ubiquitous, referring to interactions involving no fewer than three nodes \cite{battistonPhysicsHigherorderInteractions2021, zhang2023higher}. These interaction cannot be decoupled into a linear combination of pairwise interactions \cite{milo2002network, spirin2003protein,bensonHigherorderOrganizationComplex2016a}. For instance, in brain networks, cliques and cavities formed by brain regions serve as representative structures for local information processing (such as memory and computation) and global information integration, enabling the efficient operation of the brain in a distributed and parallel manner \cite{sizemore2018cliques}. Clique topologies at the level of neurons also play a crucial role in behaviors such as movement and sleep \cite{giustiCliqueTopologyReveals2015}. In protein-protein interaction networks, collaboration among multiple proteins to fulfill specific functions is prevalent \cite{lu2023more, marsh2015structure}. In ecological networks, communities formed by multiple species through predation, dependence, or mutualistic relationships constitute the fundamental building blocks of ecosystems \cite{baireyHighorderSpeciesInteractions2016,mayfieldHigherorderInteractionsCapture2017}. In social systems, whether in social relationships or communication, emergence in the form of modules or communities is common \cite{pattison2002neighborhood, alvarez2021evolutionary, majhi2022dynamics }. In engineering and technological networks, the principle of modular functional organization is widespread, with more complex tasks achieved through their cooperative efforts \cite{myersSoftwareSystemsComplex2003 }. Conversely, in networks lacking such high-order interactions, such as cycle-free networks, complexity is challenging to emerge, even with a large scale \cite{kavitha2009cycle,fan2021characterizing,zhang2021characteristics}.

Therefore, in network dismantling, merely focusing on connectivity is insufficient. It is essential to simultaneously target the disruption of connectivity and the destruction of high-order structures, prioritizing the dismantling of higher-order structures for a more practical and thorough network dismantling. We refer to this as network higher-order structure dismantling (NHSD). The NHSD represents a more generalized dismantling problem, differing not only from conventional dismantling but also from $k$-core dismantling \cite{zdeborova2016fast}, which is essentially based on $k$-core percolation (also known as bootstrap percolation) \cite{goltsev2006k, dorogovtsev2006k}. In the thermodynamic context, core percolation is exclusively concerned with the size and state of the largest unique $k$-core. However, in the context of complex networks, especially those characterized by significant community structures, distributed, or modular functionalities, efficient dismantling necessitates not only the disruption of the largest $k$-core's size and its high-order structures but also ensures the destruction of high-order structures in all other branches.

In addition to thoroughly paralyzing the structure and functionality of networks, the NHSD is closely linked to several studies, including blocking high-order random walks \cite{schaub2020random}, disrupting high-order synchronization \cite{vega2004fitness}, hindering high-order spreading \cite{iacopini2019simplicial}, and obstructing $k$-core percolation or $k$-clique percolation \cite{derenyi2005clique}, among others.

To address the challenge of higher-order structure dismantling, this paper formalizes this problem and introduced a methodology to address it. We discuss the mathematical relationship between various higher-order structures and $k$-cores, proposing to utilize the proportion of nodes in different $k$-cores as an indicator for the extent of disruption in corresponding high-order structures in the dismantling process. Subsequently, taking the scenario of edge dismantling as an example, we present the Belief Propagation-guided High-order Dismantling (BPHD) algorithm, grounded in the belief propagation model \cite{zhouSpinGlassApproach2013, mugishaIdentifyingOptimalTargets2016}. This algorithm maximizes the dismantling of high-order structures and the GCC simultaneously, with minimal cost, and ensures that the survival quantity of structures with higher orders is strictly lower than those with lower orders, not only in the GCC but also in any arbitrary branch. Experimental results demonstrate the superiority of our method over state-of-the-art benchmarks. Additionally, our algorithm exhibits significant explosive vulnerability characteristics \cite{pengUnveilingExplosiveVulnerability2023}, implying systemic fragility in the system's high-order interactions.

\section{Network higher-order structure dismantling}

\subsection{Definition}

\begin{figure}[htb]
    \centering
    \includegraphics[width=0.55\textwidth]{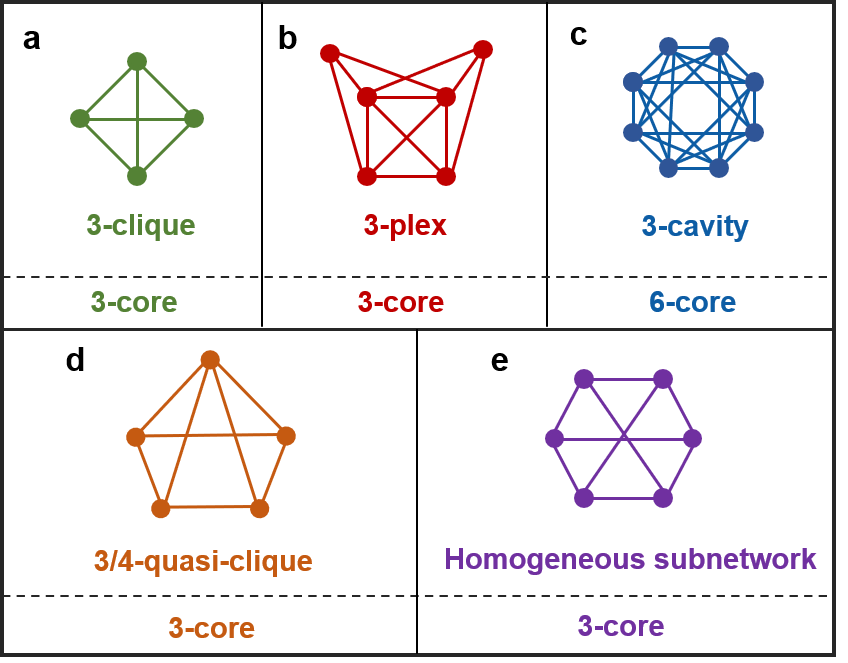}
    \caption{\textbf{Various higher-order structures and their relationships with corresponding $k$-cores.} \textbf{a}, A 3-clique, where a $k$-clique is a fully connected subgraph with $k+1$ nodes. \textbf{b}, A 3-plex, where a $k$-plex is a relaxed clique composed of $m$ nodes; the degree of any node is at least $m-k$. \textbf{c}, The smallest 3-cavity, where each node has a degree of 6. \textbf{d}, A 3/4-quasi-clique, where a $\gamma$-quasi-clique is a relaxed clique, and all nodes in it have a degree of at least $\gamma \cdot (m-1)$, with $m$ being the number of nodes, and $\gamma \in (0,1]$. \textbf{e}, A homogeneous subnetwork, where all nodes have the same degree, the same node girth, and the same node path-sum \cite{shi2013searching}. A $k$-clique, the smallest $k/2$-cavity, and homogeneous subnetwork with nodes of degree $k$ are all examples of a $k$-core; a $k$-plex is an $m-k$-core, and a $\gamma$-quasi-clique is a $\gamma \cdot (m-1)$-core.}
    \label{fig:HS}
\end{figure}

High-order structures exhibit remarkable diversity, encompassing well-known entities such as $k$-cliques and $k$-cavities \cite{sizemore2018cliques, shi2021computing}, a multitude of relaxed cliques and quasi-cliques \cite{krebs2002mapping,spirin2003protein, sanei2018enumerating,balasundaram2011clique}, homogeneous subnetworks \cite{shi2013searching},functional motifs \cite{yeger2004network, alon2007network}, modules \cite{rives2003modular}, and subgraphs with specific structures and functions \cite{bettencourt2008identification,itzkovitz2005subgraphs}, among other categories. Figure~\ref{fig:HS} illustrates these diverse structures. For a node, its order is defined as the order of the maximal higher-order structure it belongs to. For example, if a node is part of a maximal clique that is a 4-clique with 5 nodes, then its order is 4.

Taking edge dismantling as an example, it represents a more general scenario than node dismantling. The objective of the NHSD is to remove a minimum fraction $q_c$ of edges, such that it maximally disrupts both the higher-order structures and connectivity of the network. Its formal representation is given by the form:

\begin{equation}
    q_c = {\rm min} \begin{Bmatrix} 
    q \in (0,1]: S_{k\mbox{-}{\rm order}}(q) \leq H_k, S_{(k-1)\mbox{-}{\rm order}}(q) \leq H_{k-1}, \cdots, S_{2\mbox{-}{\rm order}}(q) \leq H_{2} \; {\rm and} \; S_{GCC} \leq C
    \end{Bmatrix}
\label{eq:def_NHSD_1}
\end{equation}

\noindent where $S_{k\mbox{-}{\rm order}}(q)$ denotes the proportion of nodes with an order $k$ in the residual network after removing edges with a fraction $q$. Note that this calculation includes nodes regardless of whether they are part of the GCC or not. $S_{GCC}$ denotes the number of nodes in the GCC obtained after removing edges with a fraction $q$. The constant sequences $H_k, H_{k-1}, \ldots, H_2$ correspond successively to the dismantling targets for higher-order structures with orders $k, k-1, \ldots, 2$, while the constant $C$ represents the dismantling target for connectivity. In this context, we assume that structures with higher orders $k$ are more crucial for dynamics, and therefore, they should be prioritized for disruption. This is manifested by $H_k, H_{k-1}, \ldots, H_2$ being monotonically non-decreasing as the order $k$ decreases. In general, the choice of the connectivity goal $C$ should ensure that all higher-order goals $\{S_{k\mbox{-}{\rm order}}(q) \leq H_k\}$ holds true.

\subsection{Evaluation metrics}

To evaluate the effectiveness of higher-order structure dismantling, individually counting all higher-order structures would be intricate, inefficient, and sometimes unfeasible. For instance, certain functionalities might be realized by groups of nodes with unknown precise structures or structures that deviate from typical higher-order formations. However, regardless of the specific type, these structures inherently exhibit high local connectivity, a crucial characteristic reflected by their respective $k$-core values. 

Specifically, structures like $k$-cliques or homogeneous subnetworks with a degree $k$ all belong to the category of $k$-cores; a $k$-plex composed of $m$ nodes corresponds to a $(m-k)$-core; a smallest $k$-cavity corresponds to a $2k$-core; a $\gamma$-quasi-clique composed of $m$ nodes corresponds to $\gamma \cdot (m-1)$-core, and so forth. Their relationships with $k$-cores are illustrated in Figure~\ref{fig:HS}. Additionally, motifs, modules, and specific functional subgraphs in a given network typically exhibit identifiable structures that readily reveal their associations with $k$-cores. In essence, the $k$-core can be regarded as a relaxed version of these higher-order structures at the corresponding order, as $k$ increases, the order of various higher-order structures within $k$-cores also rises. In fact, it is the existence of these higher-order structures that gives rise to the emergence of $k$-cores in the network. Therefore, by measuring the changes in the size of $k$-cores corresponding to various values of $k$, we can accurately assess the extent to which higher-order structures have been dismantled.

Therefore, a concise yet highly efficient approach is to count the node sizes of $k$-cores for each order in all branches, offering a robust reflection of the algorithm's effectiveness in dismantling higher-order structures with various levels of complexity. Consequently, Equation \ref{eq:def_NHSD_1} can be transformed into:

\begin{equation}
    q_c = {\rm min} \begin{Bmatrix} 
    q \in (0,1]: S_{k\mbox{-}{\rm core}}(q) \leq H_k, S_{(k-1)\mbox{-}{\rm core}}(q) \leq H_{k-1}, \cdots, S_{2\mbox{-}{\rm core}}(q) \leq H_{2} \; {\rm and} \; S_{GCC} \leq C
    \end{Bmatrix}
\label{eq:def_NHSD_2}
\end{equation}

\noindent where $S_{k\mbox{-}{\rm core}}(q)$ denotes the proportion of nodes within $k$-cores in the residual network after removing edges with a fraction $q$. This kind of evaluation enables us to make a reliable estimate of the quantity of higher-order structures and their response to the dismantling algorithm without explicitly calculating the specific higher-order structures.

In extreme cases, we may require the dismantling of all higher-order structures with $k \geq 2$, i.e., $H_k = H_{k-1} = \ldots = H_2 = 0$. For simplicity, $H_k, H_{k-1}, \ldots, H_2$ can be abbreviated as $H$, and $H = 0$. In this case, Equation \ref{eq:def_NHSD_2} can be written as:

\begin{equation}
    q_c = {\rm min} \begin{Bmatrix} 
    q \in (0,1]: S_{k\mbox{-}{\rm core}}(q) =  S_{(k-1)\mbox{-}{\rm core}}(q) =, \cdots, = S_{2\mbox{-}{\rm core}}(q) =  H \; {\rm and} \; S_{GCC} \leq C
    \end{Bmatrix}
\label{eq:def_NHSD_simp_2}
\end{equation}

In this paper, we focus on the dismantling of all higher-order structures, setting $H=0$ and $C=0.01N$, where $N$ represents the network size.

\section{Belief Propagation-guided High-order Dismantling}

Here, we propose a solution, the Belief Propagation-guided High-order Dismantling (BPDH) algorithm, to address the NHSD problem in the context of edge removal. The inspiration for the BPDH algorithm comes from the BPD algorithm \cite{mugishaIdentifyingOptimalTargets2016}. In scenarios involving node attacks on the GCC of the network, BPD has demonstrated the capability to induce a catastrophic collapse of the GCC in the late stages of the attack. This can be attributed to the intrinsic nature of BPD as an algorithm for constructing a minimum feedback vertex set, which aims to find the smallest set of nodes that includes at least one node from each cycle in the network \cite{zhouSpinGlassApproach2013}. Consequently, when nodes selected by BPD are removed from the network, the network transforms into a forest without cycles, leading to an explosively disruptive impact on its scale. This aligns precisely with the objectives of high-order dismantling. On the other hand, the outstanding performance of BPD has also been demonstrated through the replica-symmetric mean field theory of the spin-glass model \cite{zhouSpinGlassApproach2013} and derived results \cite{bau2002decycling}.

\begin{figure}[htb]
    \centering
    \includegraphics[width=0.65\textwidth]{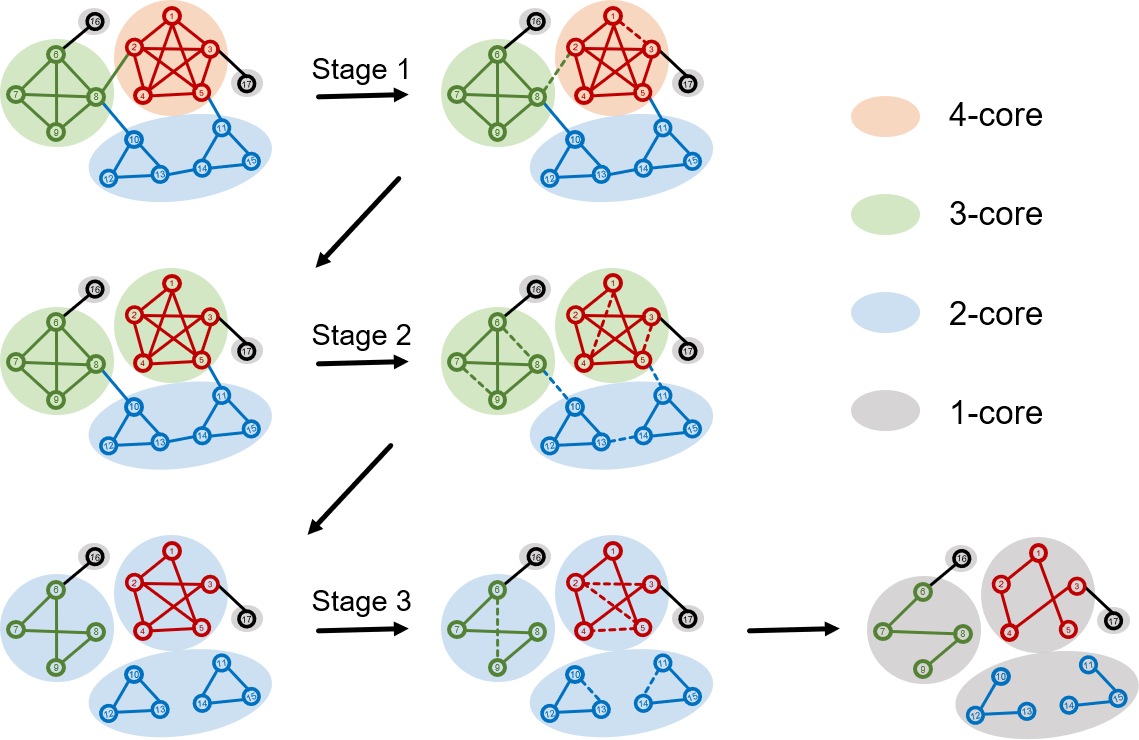}
    \caption{\textbf{The schematic representation of the BPDH algorithm in the process of dismantling higher-order structures.} Here, cliques with varying orders are used to represent different higher-order structures. Nodes of different colors indicate their membership in cliques of different orders in the original network, and shadows of different colors represent the current $k$-cores with various orders. Dashed edges depict the edges removed by BPDH in the current stage. In Stage 1, a 4-clique is dismantled into a 3-clique, resulting in the disappearance of a 4-core. In Stage 2, two 3-cliques are dismantled, leading to the loss of two 3-cores. In Stage 3, all 2-cliques and cycles are dismantled, ultimately yielding a forest.}
    \label{fig:example}
\end{figure}

We generalize the BPD algorithm to the edge dismantling scenario, resulting in the BPHD algorithm. In contrast to node dismantling, where each iteration involves removing all edges adjacent to the target node, edge dismantling, by independently removing one edge at a time, demonstrates greater generality, lower cost, feasibility, and broader applicability. Let the marginal probability $p_i^0$, similar to the case of nodes \cite{mugishaIdentifyingOptimalTargets2016}, denote the probability that edge $i$ should be prioritized for removal in each iteration, it is determined by the following expression:

\begin{equation}
    p_i^0=\frac{1}{1+e^x\left[1+\sum_{k \in \partial i} \frac{\left(1-p_{k \rightarrow i}^0\right)}{p_{k \rightarrow i}^0+p_{k \rightarrow i}^k}\right] \prod_{j \in \partial i}\left[p_{j \rightarrow i}^0+p_{j \rightarrow i}^j\right]},
    \label{eq:q0}
\end{equation}

\noindent where $x$ is an adjustable reweighting parameter and $\partial i$ denotes edge $i$'s set of neighboring edges. $p_{j \rightarrow i}^0$ and $p_{j \rightarrow i}^j$ represent the probability that edge $j$ is suitable for removal after the removal of edge $i$ and the probability that edge $j$ is suitable to be the root edge of a tree component in the absence of edge $i$, respectively. Assuming edge $i$ is removed, the two conditional probabilities are determined self-consistently through the following two belief propagation (BP) equations:

\begin{align}
p_{i \rightarrow j}^0 &=\frac{1}{z_{i \rightarrow j}}, \\
p_{i \rightarrow j}^i &=\frac{e^x \prod_{k \in \partial i \backslash j}\left[p_{k \rightarrow i}^0+p_{k \rightarrow i}^k\right]}{z_{i \rightarrow j}},
\label{eq:int_qu}
\end{align}

\noindent  where $\partial i \backslash j$ is the edge subset obtained by removing edge $j$ from set $\partial i$ and $z_{i \rightarrow j}$ is a normalization constant determined by

\begin{equation}
z_{i \rightarrow j}= 1+e^x \prod_{k \in \partial i \backslash j}\left[p_{k \rightarrow i}^0+p_{k \rightarrow i}^k\right] \times\left[1+\sum_{l \in \partial i \backslash j} \frac{\left(1-p_{l \rightarrow i}^0\right)}{p_{l \rightarrow i}^0+p_{l \rightarrow i}^l}\right] .
\label{eq:norm}
\end{equation}

Under the Bethe-Peierls approximation ~\cite{bethe1935statistical,peierls1936statistical}, we directly apply the BP equations to edges through an edge-to-node mapping, which is directly reflected in the expressions of marginal probability $p_i^0$ and the BP equations. In each iteration, BPHD removes the edge with the highest marginal probability $p_i^0$ in the current network. We employ the BPHD iteratively to remove edges until reaching the connectivity dismantling target $S_{GCC} \leq C$. At this point, we assess whether the objective of high-order dismantling has also been met. If so, the dismantling process concludes; otherwise, it continues until Equation \ref{eq:def_NHSD_simp_2} is satisfied. It is worth noting an exceptional scenario where BPHD, despite iteratively disrupting all cycles in the network, fails to satisfy $S_{GCC} \leq C$. In this case, the network transforms into a forest, and our dismantling problem degrades into a $GCC$-dismantling. Here, the strategy involves selecting edges that most rapidly reduce the size of the forest until $S_{GCC} \leq C$ is achieved. The process of dismantling higher-order structures with varying orders by BPHD in an illustrative network is depicted in Figure \ref{fig:example}. In terms of time complexity, owing to the efficiency of the Belief Propagation model, the BPHD algorithm is $O(M{\rm ln}M)$, with $M$ being the number of edges in the network.

\section{Results}

To evaluate the performance of BPHD, we conducted experiments on six networks, encompassing two classic model networks, Erdős–Rényi (ER) networks \cite{newmanNetworks2018}, and Barabási–Albert (BA) networks \cite{barabasiEmergenceScalingRandom1999}. In a BA network with a given number of nodes $N$ and model parameter $m$, indicating that each new node connected to $m$ existing nodes and all nodes in the network belong to the $m$-core. Furthermore, the presence of hub nodes facilitates the straightforward construction of higher-order structures. Additionally, four diverse empirical networks were considered: the protein-protein interaction network of yeast (Yeast) ~\cite{jeong2001lethality}, the scientific collaboration network (Collaboration) ~\cite{rossiNetworkDataRepository2015}, the email network (Email) ~\cite{rossiNetworkDataRepository2015}, and the online social network (Social) ~\cite{renGeneralizedNetworkDismantling2019}. The basic properties of these six networks are outlined in the first three columns of Table ~\ref{tb:1}.

Although the NHSD problem is a novel problem, certain existing dismantling algorithms can still serve as valuable benchmarks. Specifically, the connectivity dismantling target $S_{GCC} \leq C$ in Equation \ref{eq:def_NHSD_simp_2} aligns completely with the original objectives of Bridgeness (BG) ~\cite{chengBridgenessLocalIndex2010} and Edge Betweenness (EB) ~\cite{girvanCommunityStructureSocial2002}.  Furthermore, three node dismantling strategies renowned for their outstanding performance, namely Collective Influence (CI)~\cite{moroneInfluenceMaximizationComplex2015}, Explosive Immunization (EI)~\cite{clusellaImmunizationTargetedDestruction2016}, and Generalized Network Dismantling (GND)~\cite{renGeneralizedNetworkDismantling2019}, should also be taken into consideration.  These three algorithms, as opposed to computing scores for each edge and removing one at a time, individually calculate scores for nodes and remove all adjacent edges of the selected node each time. On the other hand, for the higher-order dismantling target in Equation \ref{eq:def_NHSD_simp_2}, we still consider BG and EB, as the removal of critical edges identified by them can inflict severe damage on higher-order structures. Additionally, we include a strategy with state-of-the-art performance in $k$-core dismantling through node removal, Cycle-Tree-Guided-Attack (CTGA) \cite{zhouCycletreeGuidedAttack2022}. The detailed definitions of these baseline methods can be found in Section \nameref{Methods}.

\begin{figure}[htb]
    \centering
    \includegraphics[width=0.8\textwidth]{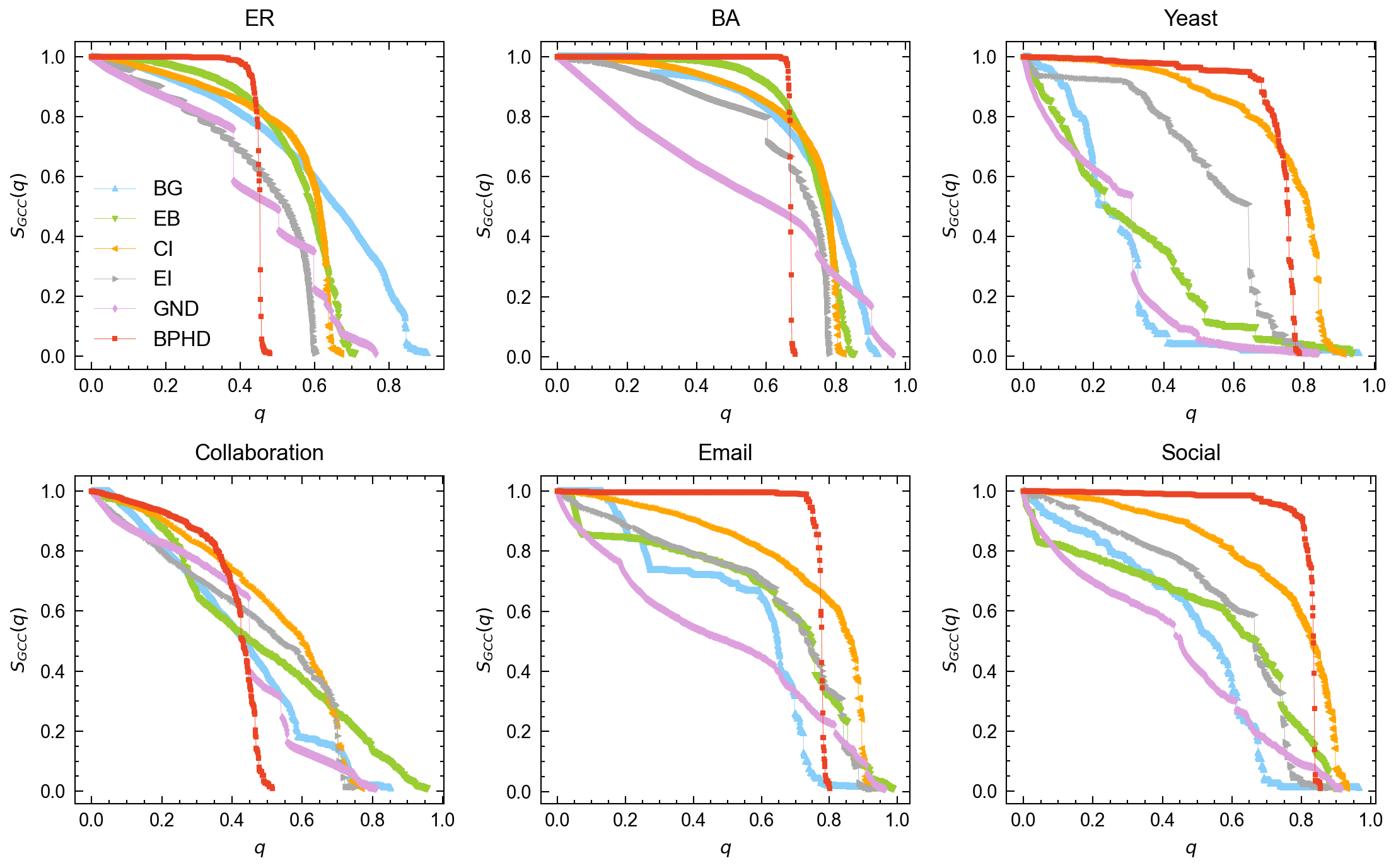}
    \caption{\textbf{Performance of BPHD in connectivity dismantling.} Here, $q$ and $S_{GCC}(q)$ represent the edge removal proportion and the corresponding relative size of GCC in the network, respectively. The dismantling objectives are set as $C=0.01N$ and $H=0$.}
    \label{fig:GCC}
\end{figure}

\subsection{Connectivity dismantling}

To facilitate clearer visualization and discussion, we analyze the performance of BPHD separately in the connectivity dismantling and higher-order dismantling. Figure \ref{fig:GCC} demonstrates that the proposed BPHD strategy outperforms baseline algorithms in GCC-dismantling, achieving the lowest-cost dismantling. The specific dismantling costs of BPHD and suboptimal methods when reaching the dismantling goal defined in Equation \ref{eq:def_NHSD_simp_2} are presented in the last two columns of Table ~\ref{tb:1}. For an ER network with an average degree of 3.5, BPHD only needs to remove a fraction of 0.48 of the edges, whereas the lowest-cost algorithm among the baselines, the EI algorithm, requires the removal of 0.6. In the case of a BA network with an average degree of 6, indicating higher density, BPHD requires the removal of 0.68, while the suboptimal algorithm necessitates 0.78. Similarly, for the four empirical networks, BPHD performs optimally in all cases. The most significant performance improvement is observed in the Collaboration network, saving 32\% of the removal cost compared to the suboptimal algorithm. In general, the higher the network density, the higher the cost.

\begin{table}[H]
\centering
\caption{The basic attributes of the six networks and the optimal and suboptimal dismantling costs in connectivity dismantling. The first four columns provide the basic attributes about the networks, including the network name, number of nodes ($N$), number of edges ($M$), and average degree ($\langle k\rangle$). The last two columns list the optimal and suboptimal dismantling costs in connectivity dismantling to achieve the dismantling targets $C=0.01N$ and $H=0$. The optimal results are indicated in bold.}
\begin{tabular}{llllcc}
\hline
              &      &       &                    & \multicolumn{2}{c}{Dismantling cost}      \\ \cline{5-6} 
Network       & $N$  & $M$   & $\langle k\rangle$ & BPHD                  & Optimal-baseline  \\ \hline
ER            & 10000 & 17500  & 3.50             & \textbf{0.48}      & 0.60           \\
BA            & 10000 & 29997  & 6.00             & \textbf{0.68}      & 0.78           \\
Yeast         & 2375 & 11693   & 9.85             & \textbf{0.79}      & 0.84           \\
Collaboration & 5094 & 7515    & 2.95             & \textbf{0.51}      & 0.75           \\
Email         & 1134 & 5451    & 9.61             & \textbf{0.80}      & 0.93           \\
Social        & 2000 & 16098   & 16.10            & \textbf{0.85}      & 0.88           \\  \hline

\end{tabular}\label{tb:1}
\end{table}

Additionally, the dismantling process of BPHD exhibits a distinctive explosive vulnerability pattern \cite{pengUnveilingExplosiveVulnerability2023}, where the early-stage GCC’s size remains nearly unchanged until reaching a certain threshold, after which it rapidly decreases. This pattern is evident across all networks except for the Collaboration network, contrasting with the continuous decline observed in benchmark algorithms. The BPHD dismantling process demonstrates a pronounced stealthiness, as the removal of crucial edges in the early stages does not significantly disrupt connectivity but focuses more on dismantling higher-order structures, as illustrated in Figure \ref{fig:example}. As the network is dismantled into a tree-like structure in the later stages, the GCC undergoes an irreversible collapse suddenly. This combination of early-stage stealthiness and late-stage abruptness poses formidable challenges to network security and robustness.

\subsection{Higher-order structure dismantling}

\begin{figure}[htb]
    \centering
    \includegraphics[width=0.8\textwidth]{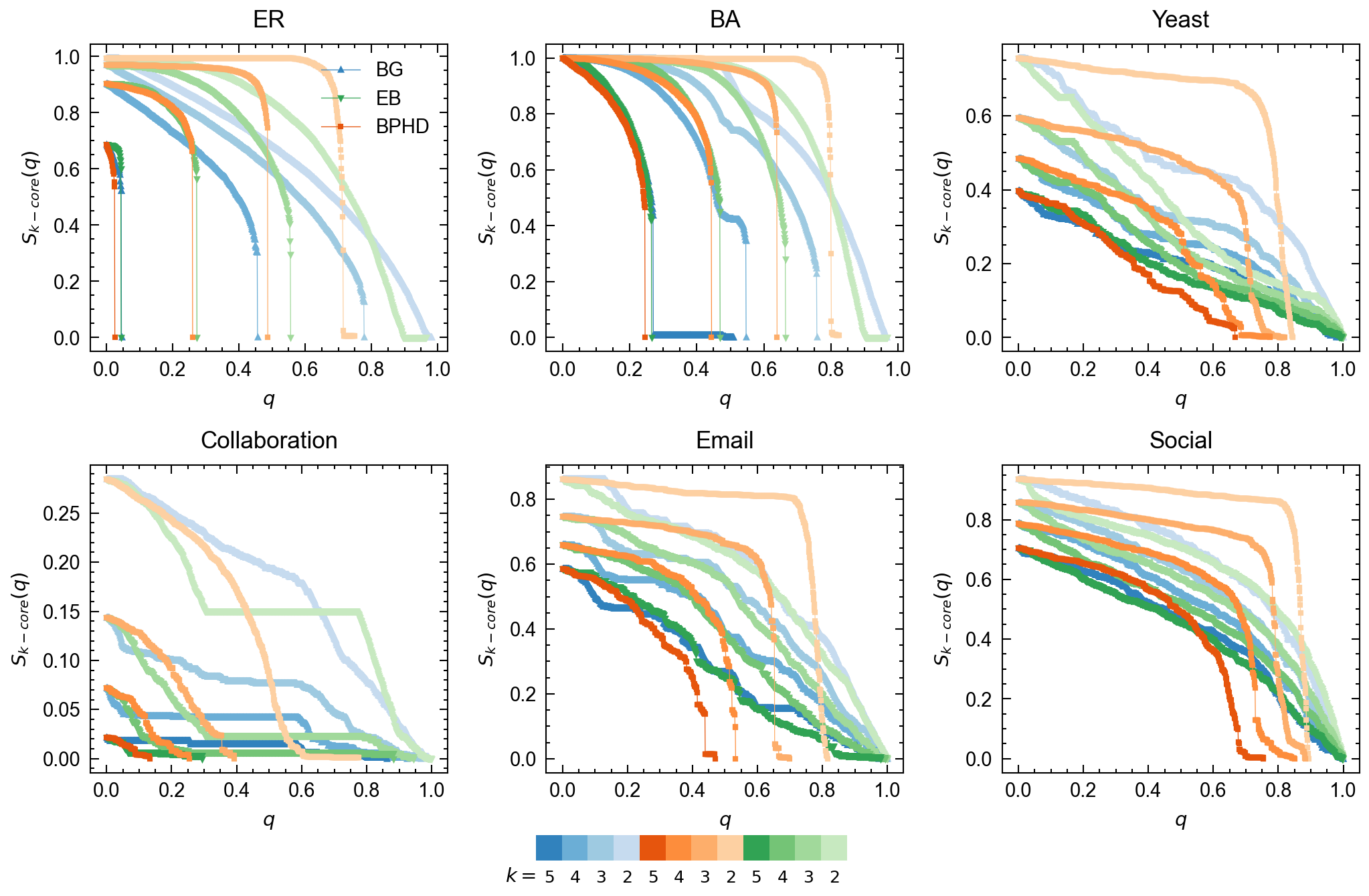}
    \caption{\textbf{Performance comparison of BPHD with two classical algorithms in higher-order structure dismantling.} Here, $q$ and $S_{k-core}(q)$ denote the edge removal proportion and the corresponding proportion of nodes with $k-core$ value equal to $k$ in all branches, respectively. Each color corresponds to a specific algorithm, with shades from dark to light indicating the dismantling results for different orders $k$. The dismantling objectives are set as $C=0.01N$ and $H=0$.}
    \label{fig:Kcore_EB_BG}
\end{figure}

\begin{figure}[H]
    \centering
    \includegraphics[width=0.8\textwidth]{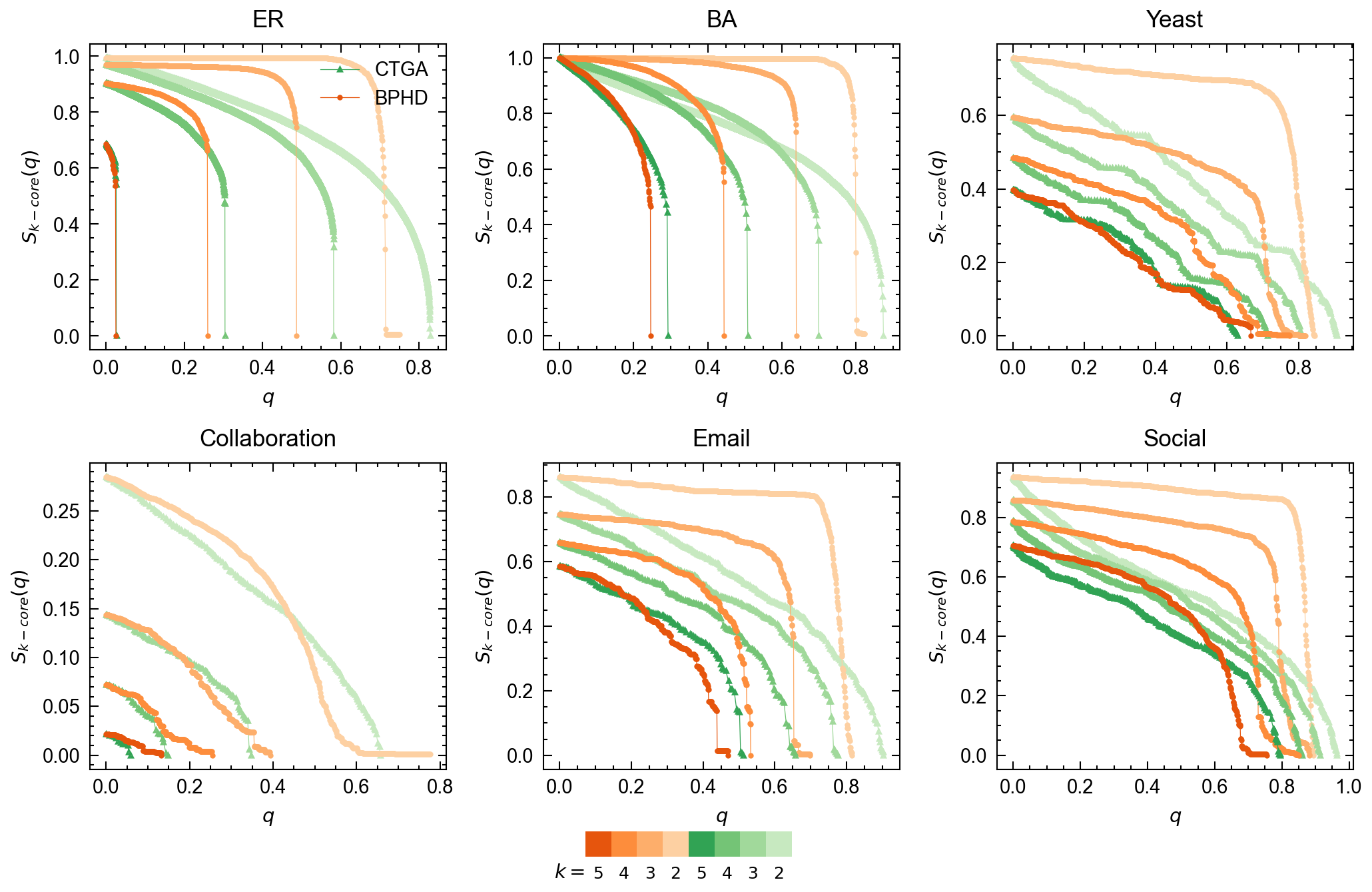}
    \caption{\textbf{Performance comparison of BPHD with the CTGA algorithm in higher-order structure dismantling.} Here, $q$ and $S_{k-core}(q)$ denote the edge removal proportion and the corresponding proportion of nodes with $k-core$ value equal to $k$ in all branches, respectively. Each color corresponds to a specific algorithm, with shades from dark to light indicating the dismantling results for different orders $k$. The dismantling objectives are set as $C=0.01N$ and $H=0$.}
    \label{fig:Kcore_real}
\end{figure}

Figures \ref{fig:Kcore_EB_BG} and  \ref{fig:Kcore_real} showcase  the results of higher-order dismantling achieved by BPHD, where we exemplify the cases of 5-core, 4-core, 3-core, and 2-core. It is noteworthy that in these results, we employed model networks with higher density, where ER and BA networks had average degrees of 7 and 10, respectively, to encompass more and higher-order structures. Consistent with Equation \ref{eq:def_NHSD_simp_2}, we utilized the relative size of $k$-cores at each order in all branches as a signal for the variation of higher-order structures of corresponding order in dismantling. Figure \ref{fig:Kcore_EB_BG} compares BPHD with two distinguished edge dismantling strategies, BG and EB. Specifically, colors ranging from dark to light correspond to orders from high to low, and all higher-order structures are sequentially eradicated by BPHD, perfectly aligning with the higher-order dismantling objective in Equation \ref{eq:def_NHSD_simp_2}, in both synthetic and empirical networks. In contrast, EB and BG do not exhibit such efficiency, as various higher-order structures persist in the final stages of dismantling, indicating their inefficacy in higher-order dismantling, especially in empirical networks. Moreover, for each order of higher-order structure dismantling, BPHD incurs the lowest dismantling cost and demonstrates a more pronounced advantage over baseline methods in empirical networks. Finally, we observe that empirical networks often require the removal of a higher proportion of edges, with the costs for extinguishing higher-order structures at different orders being closer and the discontinuity in size reduction being weaker in the later stages, indicating the stronger robustness of empirical networks compared to synthetic networks. This implies that there might be robust mechanisms at play in empirical networks that are yet unknown to us.

Figure \ref{fig:Kcore_real} illustrates the performance comparison between BPHD and CTGA. Both BPHD and CTGA are based on the belief propagation model. However, the CTGA algorithm requires setting a specific parameter $K$ for specific-order $K$-core attack tasks to achieve its optimal attack cost, whereas BPHD is parameter-free. Therefore, the $k$-core dismantling results of CTGA for orders $k=$5, 4, 3, and 2 in Figure \ref{fig:Kcore_real} correspond to the cases where $K$ is set to 5, 4, 3, and 2, respectively. In all instances, the BPHD algorithm achieves its optimal attacks against various orders of high-order structures with a single execution, a remarkable advantage compared to the CTGA algorithm. The detailed dismantling costs for BPHD and CTGA across different orders of $k$-core are presented in Table \ref{tb:2}.

We observe that BPHD outperforms CTGA in the dismantling of $k$-cores at various orders, demonstrating lower dismantling costs, except for the Yeast and Collaboration networks. This implies that the edges identified by the NPHD algorithm are genuinely important, contributing to the integrity of higher-order structures at various orders. Moreover, in terms of dismantling patterns, BPHD exhibits a more pronounced explosive vulnerability, especially at lower orders. Some of the differences between BPHD and CTGA can be attributed to CTGA's focus on node dismantling, removing all adjacent edges of the target node in each iteration. However, comparing CTGA with the results of BG and EB in Figure \ref{fig:Kcore_real}, CTGA's advantages in higher-order dismantling are evident. Combining the insights from Figures \ref{fig:Kcore_EB_BG} and \ref{fig:Kcore_real}, we arrive at a counterintuitive conclusion — higher-order structures are more fragile, requiring a minimal number of removals for effective dismantling. This contrasts with the continuous decline observed in benchmark algorithms, particularly in empirical networks depicted in Figure \ref{fig:Kcore_EB_BG}. The observed higher-order vulnerability and the covert nature of BPHD's dismantling strategy underscore the need for increased attention to the robustness of higher-order structures within a system.

\begin{table}[H]
\centering
\caption{The dismantling costs of EBPD and CTGA for higher-order structure dismantling at different orders. Specifically, each fraction represents the proportion of edges that need to be removed in the corresponding $k$-core dismantling to achieve the dismantling targets $C=0.01N$ and $H=0$. The optimal in each column results are indicated in bold.}
\begin{tabular}{lllllllll}
\hline
& \multicolumn{8}{c}{Dismantling cost}      \\
\cline{2-9}
& \multicolumn{2}{c}{2-core} & \multicolumn{2}{c}{3-core} & \multicolumn{2}{c}{4-core}     & \multicolumn{2}{c}{5-core} 
\\ \cline{2-9} 
Network       & BPHD     & CTGA            & BPHD             & CTGA    & BPHD             & CTGA             & BPHD             & CTGA            \\ \hline
ER            & \textbf{0.75} & 0.83 & \textbf{0.49} & 0.58 & \textbf{0.26} & 0.30          & \textbf{0.02}  & 0.03          \\
BA            & \textbf{0.82} & 0.87 & \textbf{0.64} & 0.70 & \textbf{0.44} & 0.51          & \textbf{0.25} & 0.29         \\
Yeast         & \textbf{0.84} & 0.91 & 0.82   & \textbf{0.81} & 0.77        & \textbf{0.71} & 0.67          & \textbf{0.63} \\
Collaboration & 0.78 & \textbf{0.66} & 0.39   & \textbf{0.35} & 0.25        & \textbf{0.15} & 0.13          & \textbf{0.06} \\
Email         & \textbf{0.82} & 0.90 & \textbf{0.70}   & 0.78 & \textbf{0.53} & 0.66        & \textbf{0.47} & 0.51         \\ 
Social        & \textbf{0.89} & 0.96 & \textbf{0.88}   & 0.91 & \textbf{0.85} & 0.86        & \textbf{0.75} & 0.80 \\ \hline
\end{tabular}\label{tb:2}
\end{table}

\section{Conclusion and discussion}

The higher-order structures within networks play a crucial role in maintaining network architecture, functionality, and giving rise to diverse complex dynamics, garnering attention across various domains. However, their vulnerability and dismantling problems, closely tied to the former, have not received sufficient consideration. In this paper, we introduce the network higher-order structure dismantling problem, building upon connectivity dismantling and specific $k$-core dismantling, and underscore the importance of completely disrupting all higher-order structures within all branches of a network. Given the diversity and, in many cases, the unknown specifics of high-order structures, monitoring the response of each individual structure to dismantling algorithms is often inefficient and, in some instances, unfeasible. Fortunately, the strong internal connectivity of most higher-order structures establishes a close association with the corresponding $k$-cores. We demonstrate this tight relationship, thereby simplifying the problem by evaluating the efficiency of higher-order dismantling algorithms through the statistical analysis of all nodes' $k$-core values. Consequently, we devise a universal framework for higher-order structure dismantling. 

Furthermore, we propose an efficient algorithm for higher-order structure dismantling, BPHD, based on belief propagation. Compared to baseline methods, BPHD achieves minimal edge removal cost, maximizing connectivity dismantling and ensuring the complete eradication of all higher-order structures of order 2 and above. It ultimately transforms the network into a forest, eliminating the possibility of sustaining any complex dynamics or functionality.

BPHD exposes an extraordinary fragility of the network core in high-order structure attacks. It achieves highly efficient dismantling of the highest-order structures with minimal cost, and remarkably, higher-order structures collapse earlier as their order increases. Additionally, both diverse high-order structures and connectivity can be explosively dismantled, manifesting as an early-stage, imperceptible process, followed by a rapid and irrecoverable collapse in the later stages. This suggests that both the core and periphery of the network exhibit explosive vulnerability, posing significant challenges for network maintenance and defense against malicious attacks.

Theoretical and practical implications of the network higher-order structure dismantling problem are vast, and it will be the focal point of our future endeavors. This approach finds applications in modeling and phase transitions in higher-order percolation \cite{zhao2022higher}, interrupting disease propagation on simplicial complex networks \cite{iacopini2019simplicial}, minimizing costs in viral marketing, communication disruption in drone swarms \cite{sharma2020communication}, as well as thorough dismantling of criminal networks and terrorist organizations \cite{xu2004fighting}.

\section*{Methods}
\label{Methods}

\textbf{Bridgeness (BG)} Bridgeness~\cite{chengBridgenessLocalIndex2010} is a local index used to measure the significance of an edge in maintaining the global connectivity of a network. For edge $e(u,v)$ , its ability to connect the large clusters in the network is defined as:

\begin{equation}
    B_{e(u, v)}=\frac{\sqrt{S_u S_v}}{S_{e(u, v)}},
    \label{eq:SI 15}
\end{equation}

where $S_u$, $S_v$ and $S_{e(u, v)}$ are the sizes of the largest clusters containing node $u$, node $v$ and edge $e(u, v)$, respectively.

\textbf{Edge Betweenness (EB)} Edge Betweenness~\cite{girvanCommunityStructureSocial2002}  is a centrality measure used to identify the edges that are most important for maintaining the global connectivity of a network. According to the idea that the more shortest paths between pairs of nodes pass through an edge $e(u, v)$, the more important the edge $e(u, v)$ is, the centrality of edge $e(u, v)$ is defined as

\begin{equation}
    EB(u, v)=\sum_{s \neq t \in V} \frac{\delta_{s t}(u, v)}{\delta_{s t}},
    \label{eq:SI 16}
\end{equation}

where $\delta_{st}$ is the number of all shortest paths between node $s$ and node $t$, and $\delta_{st}(u, v)$ is the number of all shortest paths between node $s$ and node $t$ that pass through edge $e(u, v)$. A larger EB score means greater importance of the edge.

\textbf{Collective Influence (CI)} The CI algorithm~\cite{moroneInfluenceMaximizationComplex2015} identifies the smallest set of influencers by solving the node-based optimal percolation problem. 
Given the parameter $\ell $, the value of $CI_{\ell} (u)$ of node $u$ in the network is defined as follows:

\begin{equation}
    \mathrm{CI}_{\ell}(u)=\left(k_{u}-1\right) \sum_{v \in \partial \mathrm{Ball}(u, \ell)}\left(k_{v}-1\right),
    \label{eq:SI 9}
\end{equation}

where $k_u$ is the degree of node $u$ and $\partial \mathrm{Ball} (u,\ell)$ denotes the set of nodes in the network whose shortest path length to $u$ is $\ell$.

\textbf{Explosive Immunization (EI)} The EI algorithm~\cite{clusellaImmunizationTargetedDestruction2016} is a method used to interrupt the spread of infection in a network. It combines the explosive percolation (EP) paradigm with the idea of maintaining a fragmented distribution of clusters. This algorithm heuristically utilizes two node scores, $\sigma_u^{(1)}$ and $\sigma_u^{(2)}$, to estimate a node's ability to interrupt infection propagation at two phases, which are defined as follows:

\begin{equation}
    \sigma_u^{(1)}=k_u^{(\mathrm{eff})}+\sum_{\mathcal{C}\subset \mathcal{N}_u}(\sqrt{|\mathcal{C}|}-1).
    \label{eq:SI 10}
\end{equation}

\noindent The first term $k_u^{\text {(eff) }}$ is the effective degree of node $u$, which is determined self-consistently from the original degree $k_u$:
\begin{equation}
    k_u^{(\mathrm{eff})}=k_u-L_u-M_u\left(\left\{k_v^{(\mathrm{eff})}\right\}\right),
    \label{eq:SI 11}
\end{equation}

where $L_u$ and $M_u$ are the number of leaf and hub nodes in the vicinity of $u$ respectively. During the iteration, nodes with effective degree $k_v^{(\mathrm{eff})} \geq K$ are regarded as hub nodes for a suitably chosen constant $K$. The second term is determined by the size $|\mathcal{C}|$ of cluster $\mathcal{C}$ in the set $\mathcal{N}_u$ of all clusters linked to $u$.

As the percolation process proceeds, some harmful nodes identified by $\sigma_u^{(1)}$ become harmless. To distinguish the influence of nodes more accurately, the EI algorithm uses $\sigma_u^{(2)}$ to evaluate the influence of nodes in Phase 2.

\begin{equation}
    \sigma_u^{(2)}= \begin{cases}\infty & \text { if } \mathcal{G}(q) \not \subset \mathcal{N}_u, \\ \left|\mathcal{N}_u\right| & \text { else, if } \arg \min _u\left|\mathcal{N}_u\right| \text { is unique, } \\ \left|\mathcal{N}_u\right|+\epsilon\left|\mathcal{C}_2\right| & \text { else. }\end{cases}
    \label{eq:SI 12}
\end{equation}

Here $\mathcal{G}(q)$ is the largest cluster as $qN$ nodes are removed, $|\mathcal{N}_u|$ is the number of clusters in the neighborhood of $u$, $\mathcal{C}_2$ is the second-largest cluster in $\mathcal{N}_u$, and $\epsilon$ is a small positive number (its value is not important provided $\epsilon \ll \frac{1}{N}$).
See the original article~\cite{clusellaImmunizationTargetedDestruction2016} for more details about the EI algorithm.

\textbf{Generalized Network Dismantling (GND)} The GND algorithm~\cite{renGeneralizedNetworkDismantling2019} is a method designed to fragment the network into subcritical network components with minimal removal cost. It is based on the spectral properties of the node-weighted Laplacian operator $L_w$ , and thereby transforms the generalized network dismantling problem into an integer programming problem as shown below. It achieves high performance by combining the approximation spectrum of the Laplace operator with a fine-tuning mechanism associated with the weighted vertex cover problem.

\begin{equation}
    \min_{\boldsymbol{x}=\{x_1,x_2,\dots,x_n\}} \frac{1}{4} \boldsymbol{x}^\top L_w \boldsymbol{x}
    \label{eq:SI 13}
\end{equation}

subject to

\begin{equation}
    \begin{gathered}
    \boldsymbol{1}^\top \boldsymbol{x}=0, \\
    x_i \in\{+1,-1\}, i \in\{1,2, \ldots, n\} .
    \end{gathered}
    \label{eq:SI 14}
\end{equation}

\textbf{Cycle-Tree-Guided-Attack (CTGA)} The CTGA algorithm~\cite{zhouCycletreeGuidedAttack2022}, based on the tree-packing model~\cite{zhouSpinGlassApproach2013}, extends the original model to $k$-core attack problem by allowing different tree components to be adjacent to each other and permitting additional edges within each tree component. The algorithm iteratively determines the removal probability of nodes using a coarse-grained vertex state represented in groups of four, and removes nodes one by one starting with those having the highest removal probability.

\section*{Acknowledgments}
This work is supported by the National Natural Science Foundation of China (Grant No. T2293771), the support from the STI 2030--Major Projects (2022ZD0211400) and the New Cornerstone Science Foundation through the XPLORER PRIZE. The authors express sincere gratitude to Professor Haijun Zhou for providing the CTGA algorithm code.

\section*{Author contributions}
P.P., T.F., and L.L. conceived the idea, designed the research framework, and discussed the results. P.P. collected the data, conducted the study, and drafted the initial manuscript. T.F. refined the design and experiments, and revised the manuscript. T.F. and L.L. supervised the research and performed the final manuscript editing. All authors reviewed and confirmed the methods and conclusions.

\section*{Competing interests}
The authors declare no competing interests.

\section*{Additional information}
The data and code within this work available on request from the authors

%Bibliography
\bibliographystyle{unsrt}  
\bibliography{NHD}

\end{document}